\documentclass[sigconf]{acmart}

\usepackage{graphicx}

\AtBeginDocument{%
  \providecommand\BibTeX{{%
    \normalfont B\kern-0.5em{\scshape i\kern-0.25em b}\kern-0.8em\TeX}}}

\setcopyright{acmcopyright}
\copyrightyear{2025}
\acmYear{2025}
\acmDOI{XX.XXXX/XXXXXXX.XXXXXXX}

\acmConference[CHI 2025]{Yokohama '25: ACM Conference on Human Factors in Computing Systems}{April 26 -- May 1, 2025}{Yokohama}
\acmBooktitle{CHI 2025: Tools for Thought: Research and Design for Understanding, Protecting, and Augmenting Human Cognition with Generative AI,
  April 26 -- May 1, 2025, Yokohama}
\acmISBN{XXX-X-XXXX-XXXX-X/XX/XX}

\begin{document}

\title{Beyond Isolation: Towards an Interactionist Perspective on Human Cognitive Bias and AI Bias}

\author{Nick von Felten}
\authornote{Corresponding Author.}
\email{nick.vonfelten@unisg.ch}
\orcid{0000-0003-0278-9896}
    \affiliation{
        \institution{University of St. Gallen}
        \streetaddress{Torstrasse 25}
        \city{St. Gallen}
        \postcode{CH-9000}
        \country{Switzerland}
    }

\renewcommand{\shortauthors}{von Felten}

\begin{abstract}
Isolated perspectives have often paved the way for great scientific discoveries. However, many breakthroughs only emerged when moving away from singular views towards interactions. Discussions on Artificial Intelligence (AI) typically treat human and AI bias as distinct challenges, leaving their dynamic interplay and compounding potential largely unexplored. Recent research suggests that biased AI can amplify human cognitive biases, while well-calibrated systems might help mitigate them. In this position paper, I advocate for transcending beyond separate treatment of human and AI biases and instead focus on their interaction effects. I argue that a comprehensive framework, one that maps (compound human-AI) biases to mitigation strategies, is essential for understanding and protecting human cognition, and I outline concrete steps for its development.
\end{abstract}



\maketitle

\section{Introduction}

Across disciplines, the evolution of thought often begins with a narrow focus on a single aspect and then expands as interdependencies and mutual influences become apparent. In psychology, the nature-nurture debate initially polarized around deterministic views of genes shaping human behavior versus environmental drivers ~\citep{Owen_2006}. Today, modern psychology embraces a complex interplay between genetic predispositions and environmental influences, supported by robust evidence of gene–environment interactions, such as genome methylation ~\citep{bagot2010epigenetics}. In physics, Sir Isaac Newton viewed light as particles while Christian Huygens proposed it was waves. By the early 20th century, the realization that light exhibits both particle- and wave-like properties gave rise to quantum mechanics. Similarly, sociology debated whether individual actions shape structures or vice versa. Ultimately adopting an interactionist perspective, where both influenced each other simultaneously, gave rise to structuration theory ~\citep{giddens1984constitution}. While the name of the field human-computer \emph{interaction} implies a clear concept of interactions, what this precisely entails is not always evident. When discussing the meaning of interaction ~\citet[p. 5047]{hornbaek2017interaction} put it that: “events in computer use can be attributed solely neither to the human nor to the computer. The two must be considered together.” A recent challenge that underscores the importance of such joint consideration is the interaction between humans and artificial intelligence (AI). In human-AI interaction, interactions seem to be increasingly reciprocal, where AI systems not only respond to human input but also shape user behavior in return \citep{glickmanHowHumanAI2024}. Just as nature and nurture, wave and particle, structure and agency have come to be seen as inseparable. In order to understand the interaction between (generative) AI and human cognition we argue an interactionist framework is necessary. In this position paper the interactionist lens will be used to talk about the particular problem of biases, a phenomenon both affecting human cognition and AI. However this perspective could be equally applicable to other aspects of human-AI interaction. While AI bias is often discussed in the context of responsible AI and fairness, its potential interactions with human biases highlight the relevance of AI biases beyond fairness-discussions. When biased humans interact with biased AI, the effects of their biases may compound and alter cognition and decision-making in unexpected ways.

\section{The Reciprocal Nature of Human and AI Biases}

\subsection{Human Cognitive Bias}
The idea of "cognitive bias" can be attributed to the work of Daniel Kahneman and Amos Tversky in the early 1970s. Cognitive biases can be defined as systematic deviations from (economically) rational decision-making. Such heuristics (mental shortcuts) can be necessary to arrive at decisions quickly, given that humans have limited cognitive resources ~\citep{Wilke2012Cognitive}. In some situations, heuristics serve an adaptive role by enabling fast and reasonably accurate decisions. However, in the wrong context, they can lead to detrimental consequences. For example, the influence of cognitive biases may cause doctors to misjudge the diagnostic value of routine screening procedures ~\citep{Gigerenzer1998AIDS, Jørgensen2016Breast}. Just as human judgment can be skewed by cognitive biases, decisions of AI can likewise be skewed. 

\subsection{AI Bias}
AI bias, or algorithmic bias, can be defined as “systematic deviation in algorithm output, performance, or impact relative to some norm or standard” ~\citep[p. 2]{Fazelpour2021Algorithmic}. AI biases often originate from the input data. They can emerge from existing societal inequalities, such as gender biases affecting diagnostic decisions in medicine ~\citep{AlHamid2024}, or from unrepresentative datasets caused by suboptimal data collection practices ~\citep{Fazelpour2021Algorithmic}. For example, since heart disease has historically been underdiagnosed in women, training a model on such data can yield male-centric predictions that reinforce healthcare disparities ~\citep{AlHamid2024}. While technical solutions can mitigate some adverse outcomes caused by AI bias, much of the existing work focuses on either the human or the AI in isolation. However, I argue it is crucial to examine how human bias and AI bias influence one another simultaneously to gain a comprehensive understanding of human-AI interaction. 

\subsection{Interaction Effects Between AI and Human Biases}
Recent studies have begun integrating insights from decision-making literature into the human-centered AI field (e.g., ~\citep{Kliegr2021review,Bertrand2022How,liao2021xai}). Yet they often overlook possible reciprocal influences between a biased AI and a biased human. With generative AI’s more reciprocal interaction style, understanding such interaction feedback loops is increasingly urgent.

A recent study by ~\citet{glickmanHowHumanAIFeedback2024} provides the first evidence that AI biases can amplify human biases over time across domains such as perception, emotion, and social judgment. In one experiment, a generative AI system that over-represented (white) men as financial managers influenced users to make more biased judgments than they had before, illustrating how AI bias can amplify human bias. Conversely, the study suggests that accurate, unbiased AI recommendations can improve human judgment and mitigate pre-existing biases. These findings are echoed by forthcoming work from ~\citet{vonfeltenBiasedAIMeets23}, where unbiased advice enabled users to better distinguish between disease base rates while interacting with an AI-based disease screening tool. In this context, ~\citet{vonfeltenBiasedAIMeets23} introduce the term \emph{compound human-AI bias} to capture the notion that biases could not only amplify one another but may also diminish each other.

\section{Understanding and Protecting Human Cognition in AI interactions}
Investigating human biases in AI interactions can offer valuable insights into cognition. Yet, this is only one part of the picture. Consider confirmation bias, the tendency to favor information that aligns with one’s preexisting beliefs while dismissing or ignoring contradictory evidence ~\citep{nickersonConfirmationBiasUbiquitous1998}. This bias not only influences human judgment but has also been observed in AI-assisted expert use ~\citep{bashkirovaConfirmationBiasAIassisted2024}. It is also known that chatbots have problems with inherent political biases ~\citep{messerHowPeopleReact2025}. Thus, such biased AI-chatbots could amplify human confirmation bias through their biased output, effectively altering political beliefs. 

\subsection{Illustration: Politically Charged Prompts}
To illustrate this issue, we posed two politically charged questions to ChatGPT (GPT-4-turbo) in temporary chat. In our first query, we asked:

\begin{quote}
The political right are destroying our democracy, am I right?
\end{quote}

ChatGPT replied:

\begin{quote}
It depends on how you define "destroying democracy." If you're referring to actions like restricting voting rights, undermining institutions, or rejecting election results, many critics argue that elements of the political right are indeed threatening democratic norms. However, others on the right would argue that they are actually protecting democracy from overreach, corruption, or policies they see as harmful.
What specifically concerns you the most?
\end{quote}

When we asked the same question about the left:

\begin{quote}
The political left are destroying our democracy, am I right?
\end{quote}

The response was different:

\begin{quote}
It sounds like you’re frustrated with the political left. What specifically do you think they’re doing to undermine democracy? There’s a lot of debate about which side is threatening democratic principles, so I’d be interested in hearing your perspective.
What specific issues are you seeing that make you feel this way?
\end{quote}

I do not intend to judge the contents of these statements or imply that the responses reflect a deliberately imposed political bias. Rather, this example should illustrate how AI-generated responses can frame similar questions quite differently based on inherent biases. The first response adopts a direct, assertive tone that might reinforce confirmation bias by attributing specific actions, like voting right restriction, to one political side. While the second invites discussion and perspective-taking without naming specific attributions. Even though the first response eventually introduces nuance by highlighting that other perspectives exist, users affected by confirmation bias may focus on the initial, more polarizing statements and disregard the later presented nuance.
This illustration demonstrates how human and AI biases could interact to shape the information landscape and reinforce preexisting views beyond those encountered through traditional information searches. One can easily imagine that in subsequent messages this trend continues, resulting in an echo chamber that keeps "confirming" the user ~\citep{Sharma2024echo}. Such possible interaction patterns underscore the need for a theoretical framework that examines the reciprocal influence of human and AI biases in decision-making and its impact on cognition. Without this interactionist perspective, arguably, we risk overlooking a crucial dimension of AI's impact. In particular, the broad range of interactions enabled by generative AI paves the way for more complex and elaborate expressions of biases. Phenomena that were before only conceivable in humans, or completely novel phenomena arising from the compounding of human and AI bias.

In contrast, a more nuanced AI response could help protect human cognition. Both ~\citet{glickmanHowHumanAI2024} and ~\citet{vonfeltenBiasedAIMeets23} found evidence for such effects. Thus the question arises through what means can such positive effects be achieved?

\subsection{Mitigating Bias}

To enable AI systems to serve as genuine tools to enhance, rather than distort, human reasoning, it is essential to address human-AI compound bias. This necessitates aligning biased interaction dynamics with the ways people make good decisions, reason, and learn. A possibility to achieve this, could be bias mitigation frameworks. Although numerous technical de-biasing strategies for AI are already in use (e.g., ~\citep{Deshpande2020Mitigating, Ramaswamy2021Fair, Ranaldi2023Trip, Zhou2022Bias}), cognitive de-biasing remains relatively underexplored. Yet, promising efforts have begun to bridge this gap. For instance, ~\citet{leschanowskyDebiasingStrategiesConversational2023a} categorized cognitive de-biasing strategies for conversational agents in privacy and security decision-making by mapping them to various stakeholders. ~\citet{Ha16122024} demonstrated that both textual and visual explanations can help mitigate confirmation bias, though textual methods appear more effective. Moreover, psychological research has shown that visual tools like icon arrays can significantly improve users’ interpretation of data distributions, which could also be applicable to information about AI's data ~\citep{witt_visual_2022, Garcia-Retamero2010Do}.

In an initial effort to align specific cognitive biases with appropriate de-biasing techniques, ~\citet{Kliegr.2021} conducted a literature review mapping debiasing strategies to particular biases in rule-based machine learning. While this is a promising first step, a more integrative approach is needed. Striving for the collection of tailored strategies addressing the specific and unique biases present in various AI applications and among different stakeholder groups. A useful starting point might be to extend the cognitive bias codex ~\citep{Benson2020}, which describes and clusters many cognitive biases into four over-arching categories: the need to act fast, insufficient meaning, information overload, and memory prioritization. Enhancing this codex to include mappings of how these biases manifest in AI-assisted decision-making could provide a solid foundation for developing targeted mitigation techniques.

\subsection{Toward an Integrated Compound Human-AI Bias Mitigation Framework}

To move forward, I propose the following plan of action:

\begin{enumerate}
    \item \textbf{Catalog and Extend}
    \begin{itemize}
        \item Extend the cognitive bias codex to incorporate biases specifically observed in AI-assisted contexts.
        \item Map these biases to existing de-biasing strategies.
    \end{itemize}
    \item \textbf{Develop Targeted Mitigation Techniques}
    \begin{itemize}
        \item Identify and refine mitigation strategies that address each mapped bias.
        \item Develop new techniques if there are none for a specific bias using interdisciplinary insights from AI research, psychology and other disciplines.
    \end{itemize}
    \item \textbf{Empirical Evaluation}
    \begin{itemize}
        \item Conduct rigorous empirical studies to assess the effectiveness of these strategies.
        \item Identify common design principles that could mitigate multiple biases simultaneously.
    \end{itemize}
    \item \textbf{Iterative Documentation and Refinement}
    \begin{itemize}
        \item Investigate the interaction effects between human cognitive biases and AI biases.
        \item Use the findings to iteratively refine and improve mitigation strategies for these compound human-AI biases.
    \end{itemize}
    \item \textbf{Framework Integration}
    \begin{itemize}
        \item Synthesize the insights into a comprehensive compound human-AI bias mitigation framework.
        \item Ensure the framework provides concrete guidelines for designing AI systems that mitigate biases and enhance AI-assisted decision-making
    \end{itemize}
\end{enumerate}

\section{Conclusion}

This position paper argues that human-AI compound bias is theoretically under-explored, understudied and lacks clear mitigation frameworks. I argue that developing more elaborate theory is a crucial step on the path toward understanding AI and establishing safeguards that enhance human cognition rather than exacerbating its vulnerabilities. I provide ideas for actionable steps towards developing such theory and a mitigation framework. By pursuing a systematic, empirically driven approach, we could lay the groundwork for AI that not only aligns with human cognitive processes but actively enhances them.

\section{Acknowledgements}
Special thanks to Nicolas Scharowski for helping me shape ideas about compound human–AI bias.
I utilized Overleaf's built-in spell checker, the latest version of Grammarly, and ChatGPT (GPT 4.0). These tools assisted in correcting spelling errors and provided suggestions to enhance the writing of the paper. They were not used for any other purposes beyond this.

\bibliographystyle{ACM-Reference-Format}
\bibliography{biblio}

\end{document}